\documentclass[pra,amsmath,amssymb,twocolumn]{revtex4-2}

\usepackage{dsfont}

\usepackage{graphicx}
\usepackage{chemarrow}
\usepackage{amsthm,amssymb,amsmath}
\usepackage{array}
\usepackage{hyperref}
\usepackage{xcolor}
\usepackage{bm,times,enumitem}
\hypersetup{colorlinks=true, urlcolor={black}, linkcolor={blue!80!black}, citecolor={blue!80!black}}

\begin{document}

\title{Optimizing mixing in the Rudner-Levitov lattice}

\author{I. Peshko$^{1}$, M. Antsukh$^{2}$, D. Novitsky$^{1}$, D. Mogilevtsev$^{1}$}
\address{B. I. Stepanov Institute of Physics, NAS of Belarus, Nezavisimosti ave. 68, 220072 Minsk, Belarus, \\
$^{2}$Physics Department, Belarusian State University, Minsk, Belarus}

\date{August 2023}

\begin{abstract}
Here we discuss optimization of mixing in finite linear and circular Rudner-Levitov lattices (Su–Schrieffer–Heeger lattices with a dissipative sublattice). We show that presence of exceptional points in the systems spectra can lead to drastically different scaling of the mixing time with the number of lattice nodes, varying from quadratic to the logarithmic one. When operating in the region between the maximal and minimal exceptional points, it is always possible to restore the logarithmic scaling by choosing the initial state of the chain. Moreover, for the same localized initial state and values of parameters, a longer lattice might mix much faster than the shorter one. Also we demonstrate that an asymmetric circular Rudner-Levitov lattice can preserve logarithmic scaling of the mixing time for an arbitrary large number of lattice nodes. 
\end{abstract}

\maketitle

\section{Introduction}

Mixing in continuous quantum walks is a much discussed and exploited  topic \cite{venegas}. It finds important applications in quantum computing \cite{10.5555/3286514}. It is also a very important aspect of modeling transport phenomena in quantum systems \cite{MULKEN201137}. Unitary quantum walks do not mix in a common classical sense of approaching an asymptotic distribution, but they do mix on average: the long-time time-averaged occupation probabilities of each lattice node can be considered as the limiting distribution \cite{10.1145/380752.380758}. Decohering quantum walks were shown to mix in a classical sense and to be able to do it faster than classical counterparts \cite{PhysRevA.67.042315,kendon_2007,bigger}.  

Here we address an aspect of mixing that was not much discussed before, namely, optimizing mixing in practical devices exploiting designed loss for producing delocalized stationary states from the localized input states. For such devices (be it dissipative beam-splitters and "equalizers" \cite{PhysRevLett.89.277901,PhysRevLett.91.070402,Mogilevtsev:10,Mukherjee2017DissipativelyCW,Ke:18,PhysRevA.103.023527,Dou_2022,doi:10.1021/acsphotonics.0c01053}, asymmetric distributors and propagators \cite{PhysRevX.5.021025,doi:10.7566/JPSJ.89.044003,Huang2021,myarxiv2022}, non-classical states protectors and generators \cite{PhysRevLett.89.277901,PhysRevLett.91.070402,PhysRevLett.83.3558,PhysRevLett.86.4988,Mogilevtsev:10,2009NatPh...5..633V}) minimization of the interaction time/length, i.e., mixing time, can be crucially important for practical feasibility and integrability.

The systems with designed loss considered in the paper can be treated from the standpoint of non-Hermitian physics \cite{doi:10.1080/00018732.2021.1876991}. The most intriguing aspect of non-Hermitian systems is the possibility of the spectral degeneracies known as the exceptional points much studied recently in classical and quantum optical systems \cite{doi:10.1126/science.aar7709}. In particular, in the systems obeying parity-time ($\mathcal{PT}$) symmetry, an exceptional point has a clear physical meaning as a point of spontaneous symmetry breaking resulting in transition between $\mathcal{PT}$-symmetric and $\mathcal{PT}$-symmetry-broken phases \cite{Feng_NatPhot_2017, El-Ganainy_NatPhys_2018, Ozdemir_NatMat_2019}. Among the multitude of effects associated with the exceptional points, we mention anisotropic transmission resonances \cite{PhysRevA.85.023802, PhysRevA.101.043834}, locking of light propagation direction \cite{PhysRevB.98.125102}, the effect of coherent perfect absorption and lasing \cite{Wong_NatPhot_2016, Novitsky_2019, doi:10.1021/acsphotonics.2c00790}, loss compensation \cite{PhysRevA.104.013507}, sensing with enhanced response \cite{PhysRevLett.112.203901,sensing}, resonant energy transfer enhancement \cite{PhysRevB.106.195410}, "masking" of exceptional points by quantum interference  \cite{Longhi:18}, and so on. 

Here we  show that in the non-Hermitian lattices considered in this work exceptional points define the character of the mixing, or even its very existence. 

The main points of this work are as follows. We consider systems with the so-called "dark state", i.e., the non-vacuum stationary state. This state is the basic practically used feature of the considered systems. They are supposed to function by projecting the initial state on this "dark state", thus generating it.    
In difference with decohering walks \cite{kendon_2007}, for generic lossy quantum walks the normalized occupation probability distribution might not mix in a classical sense. We connect character of mixing with the presence of exceptional points in the systems spectrum that can drastically affect scaling of the mixing time changing it from $O(N^2)$ to $O(\log(N))$, $N$ being a number of sub-systems (nodes) of the system. So, optimization of the mixing might involve designing system parameters to be in a certain position with respect to the exceptional points.
We also show that choice of the initial states can drastically fasten the mixing and considerably extend the region of $O(\log(N))$ scaling, while still having the number of the initially excited nodes much less than the total number of nodes. Thus, optimization should include also a proper choice of the initial state. 

The outline of this work is as follows. After a brief discussion of the mixing concept in Section II, we illustrate in Section III the first point described above with the example of the simplest three-mode dissipative system, namely, a dissipative beam-splitter. This structure can be considered the shortest kind of the finite Rudner-Levitov (RL) model \cite{PhysRevLett.102.065703} having a "dark state" (or the Su–Schrieffer–Heeger (SSH) model with the second sub-lattice being lossy \cite{PhysRevLett.42.1698}). We also show how mixing and non-mixing regimes arise and how mixing time behaves in dependence on the system parameters. In Section IV, we discuss mixing time behaviour in a generic system with an exceptional point where all the imaginary parts of the eigenvalues are coalesced. In Section V, we consider a finite linear RL model and show how the mixing time depends on the position of exceptional points. Here we show how one can initially excite only few nodes to extend $O(\log(N))$ scaling. Also, we show how to exploit the structure of eigenvectors to achieve the fastest localization with the minimal number of the initially excited nodes, $M$. In Section VI, we demonstrate how the found mixing optimization can be achieved for the circular RL lattice. In particular, we show that for the same values of the interaction constants, loss rates and initial states, the circlular lattice can exhibit much faster mixing for the same number of nodes. Moreover, we  demonstrate that the asymmetric circular RL lattice  can have log-like fast mixing scaling for arbitrary $N$.   Also, we show how for the circled lattice diabolic points \cite{doi:10.1098/rspa.1984.0022} can convalesce into an exceptional point.

\section{Mixing} 

Here we clarify the concept of mixing with respect to dissipative structures. We consider systems described by the following generic equation for the vector of amplitudes $\vec{\psi}$
\begin{equation}
\frac{d}{dt}\vec{\psi}=-i{\mathbf{H}}\vec{\psi}, 
\label{gen}
  \end{equation}
where the elements of the vector $\vec{\psi}$, i.e., the amplitudes $\psi_j$, correspond to the nodes of the lattice. This lattice is described by the effective non-Hermitian Hamiltonian ${\mathbf{H}}$. Notice that generally Eq.(\ref{gen}) describes a particular case of state dynamics resulting from the master equation describing the lattice. For instance, the amplitudes $\psi_j$ might be indeed the amplitudes of coherent states propagating through the lattice of single-mode waveguides or the off-diagonal elements of the single-particle density matrix \cite{Mogilevtsev_2015}.  

As it was already mentioned in the Introduction, we consider practical systems that function as passive state filters by projecting an initial state on the particular "dark" state (which can be a non-classical and even entangled one \cite{PhysRevX.5.021025,doi:10.7566/JPSJ.89.044003,Huang2021,myarxiv2022,Mogilevtsev_2015}). Thus, the eigenvalues $\lambda_k$ of the Hamiltonian ${\mathbf{H}}$ satisfy $\mathrm{Im}\{\lambda_k\}\leq 0$, the index $k=0,1\ldots \mathrm{dim}({\mathbf{H}})-1$. For simplicity sake, we assume that the "dark" state is unique, i.e., $\lambda_0=0$, and $|\lambda_k|>0$ for $k>0$. Also, because of passivity, the system has the vacuum as the stationary state, $\psi_j=0, \quad \forall j$. 

Loss leads to the change of total probability
$P_{total}(t)=\sum\limits_{\forall j} |\psi_j(t)|^2$, which is no more equal to $1$ (notice, that this can be accompanied by existence of other integrals of motion; for examples, for coherent diffusive photonic systems considered in Refs. \cite{Mukherjee2017DissipativelyCW,Mogilevtsev_2015} one has preservation of the sum of coherencies, $\sum\limits_{\forall j} \psi_j(t)=\mathrm{const}$). Thus, to describe mixing one needs to introduce the normalized occupation probabilities
\begin{equation}
p_j(t)={|\psi_j(t)|^2}/P_{total}(t)
\underset{\scriptscriptstyle t \rightarrow \infty}{\rightarrow} p_j^{(st)},
\label{p}
\end{equation}
with $p_j^{(st)}$ being the mixed time-independent occupation distribution. Thus, the mixing time is defined in a standard classical way as \cite{aldous}
\begin{equation}
T_{mix}(\epsilon)=\min\{t\ge 0: \sum\limits_{\forall j}|p_j(t)-p^{st}_j|\le \epsilon\}
\label{mix}
\end{equation}
for  $\epsilon>0$. 
Notice that the normalization in Eq.(~\ref{p}) actually makes ambiguous the definition of the mixed distribution. If the initial state is not orthogonal to the "dark" state, the mixed distribution is unique and is defined by this "dark" state. This kind of mixing we call "conventional" throughout the paper. When the initial state is orthogonal to the "dark" state, the system eventually decays to the vacuum. However, the normalized distribution (\ref{p}) might also mix to some limiting distribution $p^{(st)}_j$. This distribution might depend on both the initial state and the parameters of the lattice. This kind of mixing we term as "unconventional". Finally, as we will see below, for the initial state orthogonal to the "dark" state, mixing might not occur at all. 

\begin{figure}[htb]
\begin{center}
\includegraphics[width=0.8\linewidth]{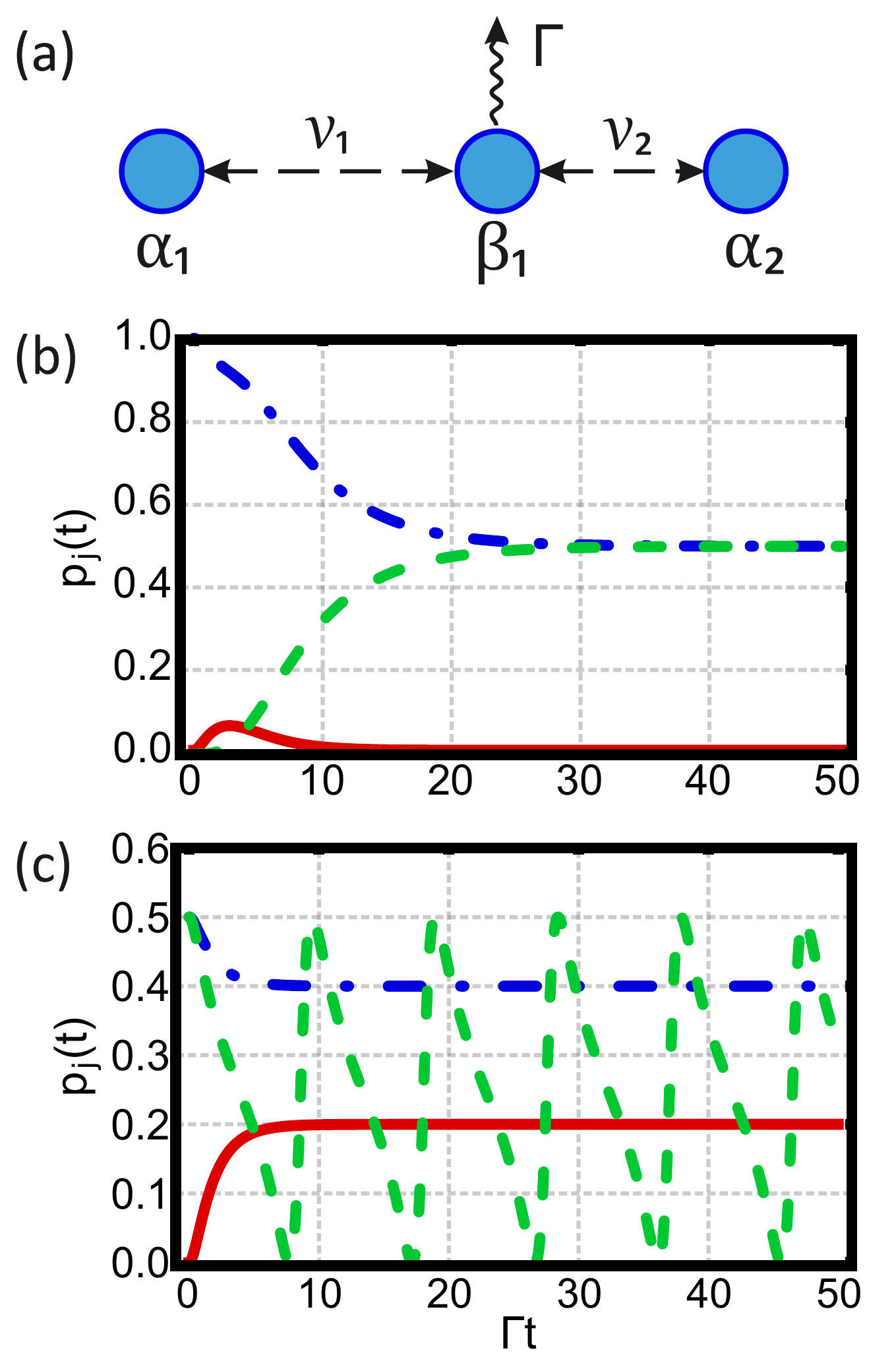} 
\end{center}
\caption{(a) A scheme of the dissipative beam-splitter (DBS) model. The nodes $\alpha_{1,2}$ are unitary coupled with the node $\beta_1$ with the coupling rates $v_{1,2}$, but not with each other; the node $\beta_1$ is subject to the loss with the rate $\Gamma$. (b-c) Illustration of the mixing regimes of the symmetric DBS described by the Hamiltonian (\ref{ham1}) with $v_1=v_2$. (b) Solid, dashed and dash-dotted lines correspond to the normalized probabilities $p_{2,3,1}(t)$ given by Eq.(\ref{p}) for the conventional mixing regime; initially only the node $\alpha_1$ is excited, $v/\Gamma=0.4$. (c) Solid and dash-dotted lines illustrate unconventional mixing for initial excitation of both the nodes $\alpha_{1,2}$ with equal amplitudes; solid and dash-dotted lines correspond to $p_{1,2}(t)$ for $v/\Gamma=0.4$. Dashed line illustrates absence of mixing for $p_1(t)$ corresponding to  $v/\Gamma=0.6$ for initial excitation of both the nodes $\alpha_{1,2}$ with equal amplitudes.}
\label{fig1}
\end{figure}

\section{Dissipative beam-splitter}

To demonstrate the essential points of our discussion, let us consider a simple structure consisting of just three unitary coupled nodes with {the corresponding amplitudes $\alpha_{1,2}$ and $\beta_1$ with the middle one corresponding to $\beta_1$ subject to loss (Fig.\ref{fig1}(a)).} This shortest case of the RL model with a "dark" state \cite{PhysRevLett.102.065703} can be realized, for example, with single-mode waveguides \cite{Eichelkraut:14,Mukherjee2017DissipativelyCW,Ke:18,PhysRevA.103.023527,Dou_2022,doi:10.1021/acsphotonics.0c01053}, with  coherences of coupled two-level systems \cite{Mogilevtsev_2015} or three-site Bose-Hubbard model \cite{PhysRevA.82.043621}. This dissipative beam-splitter (DBS)
is described by the following Hamiltonian
\begin{equation}
{\mathbf{H}}=\begin{pmatrix} 
0 & v_1 & 0 \\
v_1 & -i\Gamma & v_2 \\
0 & v_2 & 0
\end{pmatrix}
\label{ham1}   
\end{equation}
where the real $\Gamma>0$ is the loss rate of the middle node denoted as $\beta_1$ in Fig.\ref{fig1}(a). For simplicity sake, we assume real unitary coupling rates $v_{1,2}$.

Despite simplicity, the scheme described by Eq.(\ref{ham1}) contains rich physics. First of all, it can exhibit both $\mathcal{PT}$ and anti-$\mathcal{PT}$ symmetries. Indeed, for the limiting asymmetric case, when one of the coupling rates vanishes, $v_j\rightarrow 0$, the remaining part of the system (two coupled nodes) is $\mathcal{PT}$-symmetric \cite{PhysRevLett.103.093902,PhysRevLett.101.080402}. In that case it was also observed how the strong designed loss in the node $\beta_1$ leads to the effective decoupling of the remaining side node (termed "loss-induced transparency" \cite{PhysRevLett.101.080402}). In the other limit of $\sqrt{v_1^2+v_2^2}/\Gamma\rightarrow 0$, the system becomes anti-$\mathcal{PT}$-symmetric \cite{PhysRevA.96.053845,Qin:21}, since one can adiabatically exclude the middle node $\beta_1$ resulting in the so-called "dissipative coupling" \cite{PhysRevA.79.023811,PhysRevA.82.043621,Mogilevtsev:10}.

\begin{figure}[htb]
\begin{center}
\includegraphics[width=0.8\linewidth]{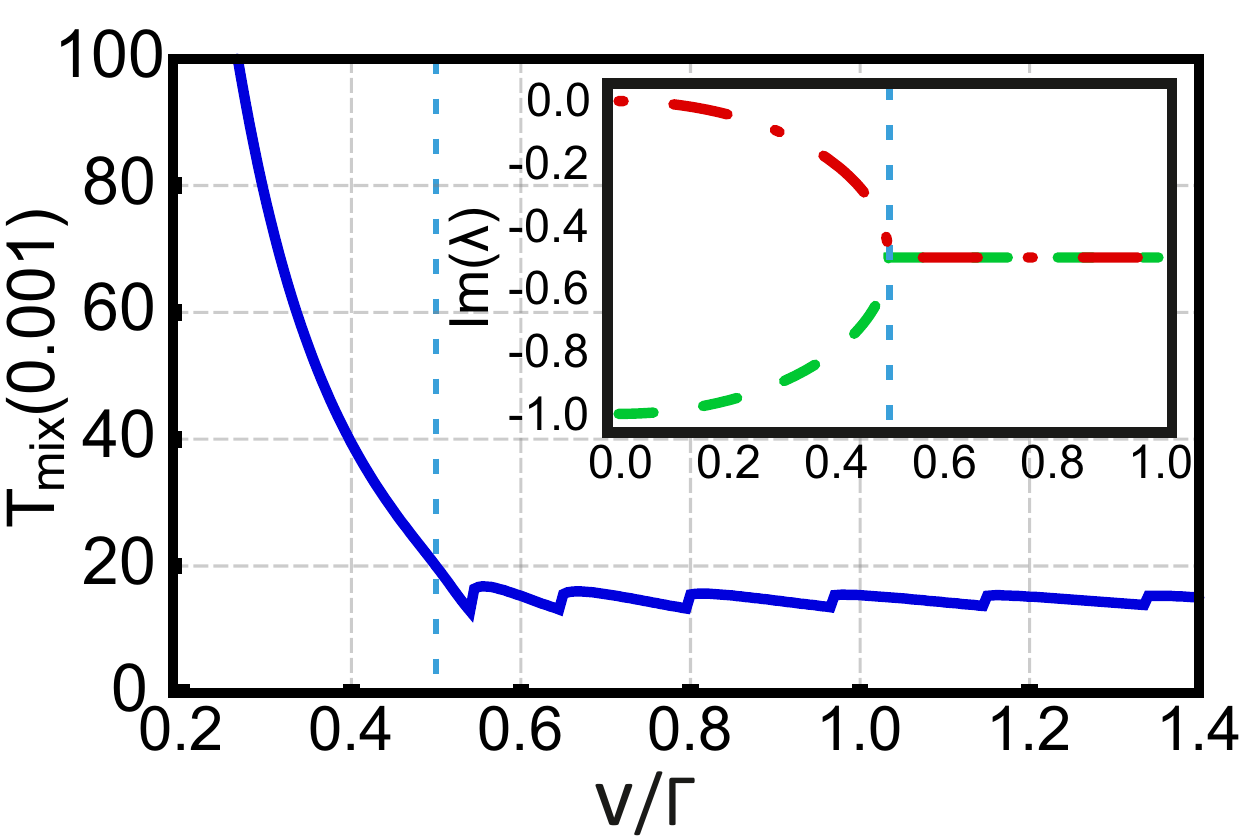} 
\end{center}
\caption{Mixing time for the symmetric DBS in the units of $\Gamma^{-1}$ for the initially excited node $\alpha_1$ as given by Eq.(\ref{mix}) with $\epsilon=0.001$ in dependence on $v/\Gamma$; the vertical line denotes the position of the exceptional point; in the inset imaginary parts of the eigenvalues of the Hamiltonian (\ref{ham1}) are shown in dependence on the ratio $v/\Gamma$. The dashed line corresponds to $\lambda^+$, the dotted line corresponds to $\lambda^-$ of Eq.(\ref{eig1}) The vertical dashed line shows the position of the exceptional point.}
\label{fig1mix}
\end{figure}

Here we show how another case of $\mathcal{PT}$-symmetry arises in this DBS for arbitrary values of $v_{1,2}$ and $\Gamma$. This symmetry is connected with the "dark" state and the mixing rate. {Indeed, let us introduce the amplitude, $v$, and phase, $\phi$, parameters characterising the coupling such that $v_1=v\cos{\phi}, \quad v_2=v\sin{\phi}$, $v=\sqrt{v_1^2+v_2^2}$, and collective variables $A$ and $\alpha$ are corresponding to the amplitudes of the "bright" and "dark" states}
\begin{equation}
A=\alpha_1\cos{\phi}+\alpha_2\sin{\phi}, \quad \alpha=\alpha_2\cos{\phi}-\alpha_1\sin{\phi}.
\label{col1}
\end{equation}
It is easy to see from Eqs.(\ref{gen}) and (\ref{ham1}) that $\alpha(t)=\mathrm{const}$. Actually, it describes projection of the initial state on the "dark" eigenstate of the Hamiltonian (\ref{ham1}) corresponding to the eigenvalue $\lambda_0=0$. The two other eigenvalues are
\begin{equation}
\lambda^{\pm}=-\frac{i}{2}(\Gamma\pm\sqrt{\Gamma^2-4v^2}).  
\label{eig1}
\end{equation}
Dynamics of the variables $A,\beta_1$ are described by the effective Hamiltonian 
\begin{equation}
{\mathbf{H}}_2=\begin{pmatrix} 
0 & v  \\
v & -i\Gamma  
\end{pmatrix},
\label{ham2}   
\end{equation}
which is $\mathcal{PT}$-symmetric. As is typical for the $\mathcal{PT}$-symmetric systems \cite{PhysRevLett.103.093902,PhysRevLett.101.080402,Longhi:18}, the spectrum (\ref{eig1}) shows an exceptional point at $v/\Gamma=1/2$, where the eigenvalues $\lambda^{\pm}$ coalesce (see the inset in Fig.\ref{fig1mix}). The position of this point with respect to the system's parameters, i.e., the ratio $v/\Gamma$, crucially affects both mixing and the character of dynamics. If projection of the initial states on the "dark" state is finite, i.e., $\alpha(0)\neq 0$, for all the finite ratios $v_{1,2}/\Gamma$ the system is always mixed conventionally with $p_2^{(st)}=0$ and $p_{1,3}^{(st)}=v_{2,1}^2/v^2$. An illustration of this behavior is given by Fig.\ref{fig1}(b) at the point before the exceptional one, $v/\Gamma=0.4$. This kind of dynamics underlies practical usability of DBSs. By loosing part of input energy, a waveguide DBS robustly reproduces outputs with needed amplitude ratios and fixed phase between them \cite{Ke:18,PhysRevA.103.023527}. 

However, nodal population dynamics becomes quite different for $\alpha(0)=0$. 
This can be understood by considering the analytic solution of the system given by Eq.(\ref{ham2}):
\begin{eqnarray}
\nonumber
\beta_1(t)=c^+e^{-i\lambda^+t}-c^-e^{-i\lambda^-t}, \\
\label{sol1}
\quad c^{\pm}=\frac{-vA(0)+(i\Gamma+\lambda^{\mp})\beta_1(0)}{\lambda^--\lambda^+}, \\
\nonumber
A(t)=\frac{i}{v}(\frac{d}{dt}\beta_1(t)+\Gamma\beta_1(t)). 
\end{eqnarray}
This solution immediately shows that beyond the exceptional point, i.e., for $v/\Gamma>1/2$, the system does not mix for $\alpha(0)=0$, so that Eqs.(\ref{sol1}) give non-decaying oscillations of the normalized occupation probabilities (see the dashed line showing the first node normalized population in Fig.\ref{fig1}(c)). Before the exceptional point, unconventional mixing still occurs (see solid and dash-dotted lines in Fig.\ref{fig1}(c)), but the values of the stationary nodal populations $p_j^{(st)}$ now become dependent on the initial excitation and the decay rate $\Gamma$. 

So, it is entirely unsurprising that conventional mixing time is very much dependent on the position of the exceptional point with respect to the system parameters. {As one can see in the inset of Fig.\ref{fig1mix} showing imaginary parts of the eigenvalues $\lambda^{\pm}$ in dependence on $v/\Gamma$,} before the exceptional point, i.e., for $v/\Gamma < 1/2$, the mixing time (\ref{mix}) is defined by $\lambda^-$ and quickly increases with $\Gamma$ (see the illustration in Fig.\ref{fig1mix} for the symmetric structure with $v_1=v_2$ and $\epsilon=0.001$). Beyond the exceptional point $T_{mix}(\epsilon)$ is defined mostly by the required precision $\epsilon$. Curiously and importantly for the practical design, the minimal mixing time is achieved in the vicinity of the exceptional point.

\begin{figure}[htb]
\begin{center}
\includegraphics[width=0.8\linewidth]{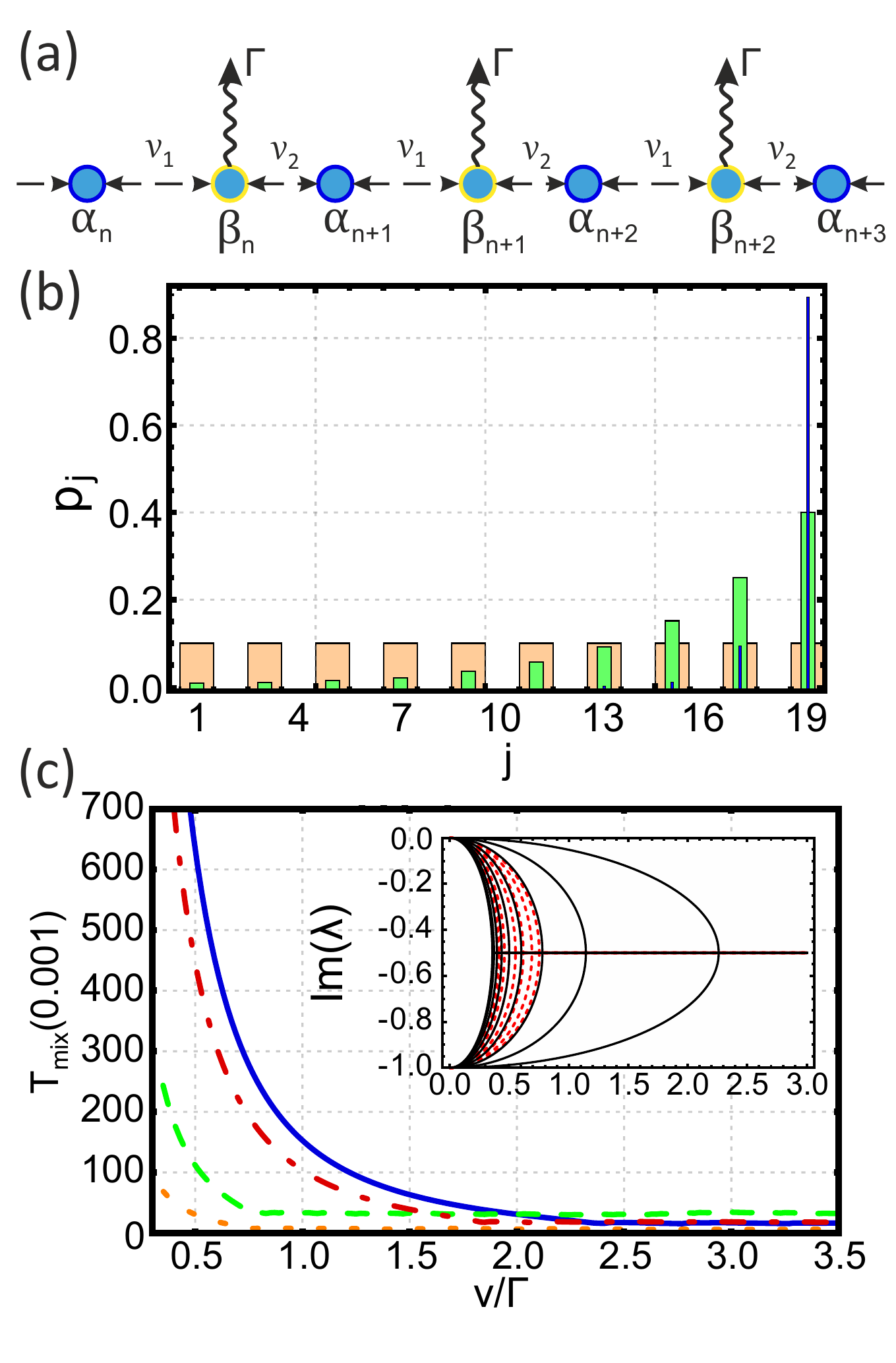} 
\end{center}
\caption{(a) A scheme of the linear RL lattice. Each lossy node $\beta_{n}$ is unitarily coupled with the node $\alpha_{n,n+1}$ with the corresponding coupling rates $v_{1,2}$, the node $\beta_n$ is subject to the loss with the rate $\Gamma$. (b) Examples of stationary distributions corresponding to a non-decaying eigenstate; thick, medium, and thin bars show $p_k^{st}$, correspondingly, for $\phi=0.25\pi,0.21\pi,0.1\pi$. (c) $T_{mix}(\epsilon)$ dependence in the units of $\Gamma^{-1}$ on the ratio $v/\Gamma$ for different $\phi$. The solid, dash-dotted and dashed curves correspond to $\phi=0.25\pi, 0.21\pi, 0.1\pi$ and a single node initially excited, $\alpha_1(0)\neq 0$; the dotted curve corresponds to $\phi=0.1\pi$ and only $\alpha_{N_L+1}(0)\neq 0$. The solid lines in the  inset show  $\mathrm{Im}(\lambda_k)$ for the symmetric case, in dependence on the ratio $v/\Gamma$, and dotted lines show $\mathrm{Im}(\lambda_k)$ for the strongly asymmetric case, $\phi=0.1\pi$. For all the panel (b), $N_L=9$.}
\label{fig2}
\end{figure}

\section{General considerations}

One can easily surmise that mixing features demonstrated by the DBS considered in the previous Section are to be observed in more complicated systems with similar spectral features. Indeed, let us have a single-state passive filtering system described by Eq.(\ref{gen}) with the effective Hamiltonian $\mathbf{H}$. The corresponding eigenequation is as follows
\begin{equation}
\mathbf{H}\vec{\phi}_n=\lambda_n\vec{\phi}_n,
\label{eig2}
\end{equation}
with eigenvalues satisfying $\lambda_0=0$, ${\rm Im}(\lambda_n)<0, \forall n\in[1,N-1]$ in all the considered range of the Hamiltonian parameters; $N$ is the total number of nodes in the system; $\vec{\phi}_n$ are eigenvectors. Now let us assume that in a certain region of the parameter space, for $\forall n\in[1,N]$, one has the coalescence of all the imaginary parts, i.e. $\lambda_n=-i\gamma+\omega_n$. Away from the exceptional spectral points, one can write a solution 
\begin{equation}
{\psi}_k=c_{k0}+e^{-\gamma t}\sum\limits_{n=1}^{N-1}c_{kn}e^{-i\omega_n t}
\label{psi1}
\end{equation}
where the coefficients $c_{kn}$ are defined by the projection of the initial state on the corresponding eigenvector $\vec{\phi}_n$. So, one has some positive $C$, so that $|c_{kn}|\leq C, \forall k,n$. From Eq.(\ref{psi1}) it is easy to estimate the conventional mixing time for the normalized distribution (\ref{p}). For conventional mixing, when $c_{k0}\neq 0$ for some $k> 0$, from  Eqs. (\ref{p}), (\ref{mix}), and (\ref{psi1}) it follows that 
\begin{equation}
\gamma T_{mix}(\epsilon)\leq O(\log\{CN/\epsilon\}).
\label{tmixlog}
\end{equation}
The estimate (\ref{tmixlog}) is rather favorable for achieving mixing in the structures with $N\gg1$. On the other side, in the absence of eigenvalues coalescence, i.e., for a set of different ${\rm Im} (\lambda_n)<0$, the conventional mixing time is defined in a standard way by a spectral gap ($\min|{\rm Im}(\lambda_n)|, n>0$). This scaling can be quite unfavorable. For example, for a classical linear Markov chain $T_{mix}$ scales as $N^2$ \cite{aldous}. 

For practical mixing structures, the question is when it is feasible to reach the coalescence region of parameters and/or generally keep the scaling (\ref{tmixlog}). Also, the question is what can one do to minimize mixing time when it is not feasible to reach the coalescence region, in particular, when only a part of the eigenvalues coalesced. 

Further we demonstrate that using symmetry of the eigenvectors for partially coalesced spectrum, it is still possible to restore scaling (\ref{tmixlog}) while keeping the mixing device practically useful, i.e., keeping the number of the initially excited nodes reasonably small and also keeping the set of these nodes well localized. We demonstrate it on the example of two very common workhorse models of dissipative quantum walks: linear and circular Rudner-Levitov lattices \cite{PhysRevLett.102.065703}.

\section{Linear RL lattice}

Now let us consider how mixing occurs in a particular example of the generic structures considered in the previous Section: the finite Rudner-Levitov (RL) linear lattice \cite{PhysRevLett.102.065703} with $N_L$ lossy nodes and $N_L+1$ lossless ones. The scheme of such lattice is shown in Fig.\ref{fig2}(a), where the amplitudes corresponding to the nodes of the lossless sublattice are termed as $\alpha_n$ and the amplitudes corresponding to the nodes of the lossy sublattice are termed as $\beta_n$. The effective Hamiltonian of the considered structure is a tridiagonal matrix with the following non-zero elements
\begin{eqnarray}
\nonumber
H_{2n,2n}=-i\Gamma, \quad H_{2n-1,2n}=H_{2n,2n-1}=v_1,\\
H_{2n,2n+1}=H_{2n+1,2n}=v_2, \quad n=1,2\ldots N_L,
\label{ham3}
\end{eqnarray}
where we similarly to the DBS case parameterize the unitary coupling rates (taken for simplicity to be real) as $v_1=v\cos{\phi}, \quad v_2=v\sin{\phi}$, $v=\sqrt{v_1^2+v_2^2}$. The RL lattice described by the Hamiltonian (\ref{ham3}) is a simple extension of the DBS (\ref{ham1}) retaining its essential functional feature: a single non-vacuum stationary state allowing for conventional mixing. Further, we will concentrate on this type of mixing.

\subsection{Mixing and asymmetry}
First of all, let us highlight the role of {unequal coupling  (i.e., $v_1\neq v_2$) for the considered RL lattice devices (farther in the text we term such unequal coupling as "asymmetry" in coupling).} From the Hamiltonian (\ref{ham3}), it is obvious that the ratio of the nodal amplitudes $\alpha_n/\alpha_{n+1}$ in the stationary "dark" state is $v_2/v_1$. So, asymmetry of the coupling leads to the asymmetry of the "dark" state population distribution and might severely limit functionality of the device aimed at the distribution of population among the nodes. An example for the RL lattice with 19 nodes can be seen in  Fig.\ref{fig2}(b): The thick, medium, and thin bars show $p_k^{st}$ for symmetric ($\phi=\pi/4$), slightly asymmetric ($\phi=0.21\pi$), and strongly asymmetric cases ($\phi=0.1\pi$), respectively. For $\phi=0.1\pi$ the "dark" state is strongly localized near the  edge of the lattice, so that the initial state would effectively mix only through a few edge nodes.

As we have shown in the previous Section, asymmetry does not affect the spectrum of the DBS (\ref{eig1}). It is not so for the RL lattice with larger number of nodes. Qualitatively, the behaviour of the imaginary part of the eigenvalues, $\mathrm{Im}(\lambda_k)$, is similar for both the symmetric and asymmetric cases as one can see for the lattice with $N=19$ nodes in the inset of Fig.\ref{fig2}(c): The solid lines show $\mathrm{Im}(\lambda_k)$ as a function of the ratio $v/\Gamma$ for the symmetric case, and the dotted lines show $\mathrm{Im}(\lambda_k)$ for the strongly asymmetric case, $\phi=0.1\pi$. One can see in both cases that for sufficiently large ratio $v/\Gamma$, all the imaginary parts of eigenvalues coalesce. With lowering $v/\Gamma$, the system goes through a number of exceptional points (there are $N_L$ of them). Eventually, for sufficiently small $v/\Gamma$, the system enters the region of dissipative coupling. Both symmetric and asymmetric systems enter this region for close values of $v/\Gamma$. However, the spread of exceptional points is larger for the symmetric case. Coalescence is broken for larger $v/\Gamma$. These features are obvious from the spectra of non-zero eigenvalues easily obtained analytically due to a simple tridiagonal structure of the matrix (\ref{ham3}) \cite{toepl}:
\begin{equation}
\lambda^{\pm}_k(\phi)=-\frac{i}{2}\left(\Gamma\pm\sqrt{\Gamma^2-4v^2\mu_k(\phi)}\right),  \label{eig3}
\end{equation}
where $\mu_k(\phi)=1+\sin2\phi \cos\left\{\frac{k\pi}{N_L+1}\right\}$ and $k=1,2\ldots N_L$.

\begin{figure}[htb]
\begin{center}
\includegraphics[width=\linewidth]{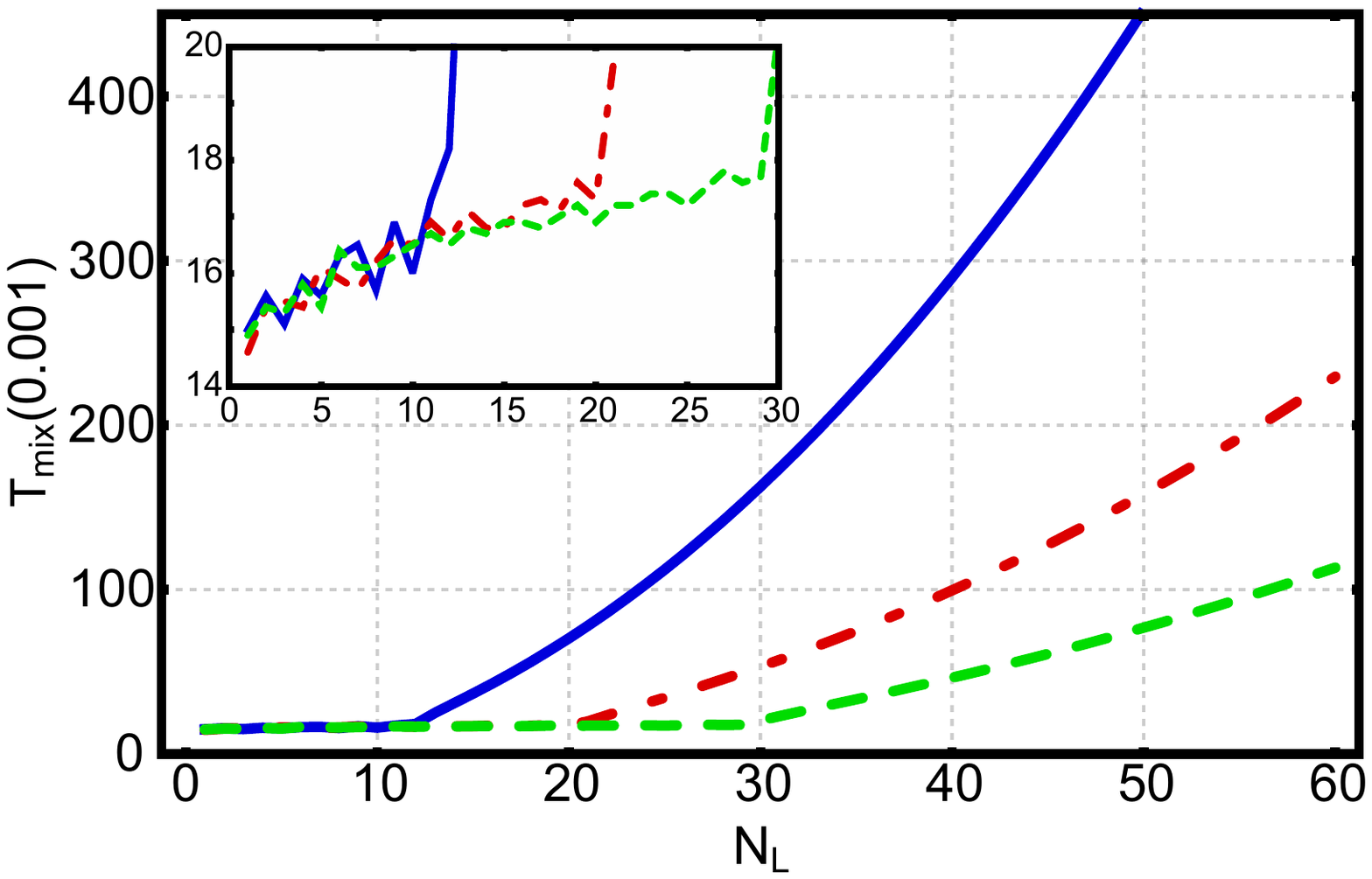} 
\end{center}
\caption{ Dependence of the mixing time $T_{mix}(0.001)$ shown in the units of $\Gamma^{-1}$ on the number of lossy nodes $N_L$ for the symmetric RL lattice. Solid, dash-dotted, and dashed lines correspond to $v/\Gamma=3,5,7$, respectively. For all the curves $\epsilon=0.001$. The inset shows $T_{mix}$ for smaller interval of $N_L$. Initially only the first losless node is excited.}
\label{fig3}
\end{figure}

\begin{figure}[htb]
\begin{center}
\includegraphics[width=0.9\linewidth]{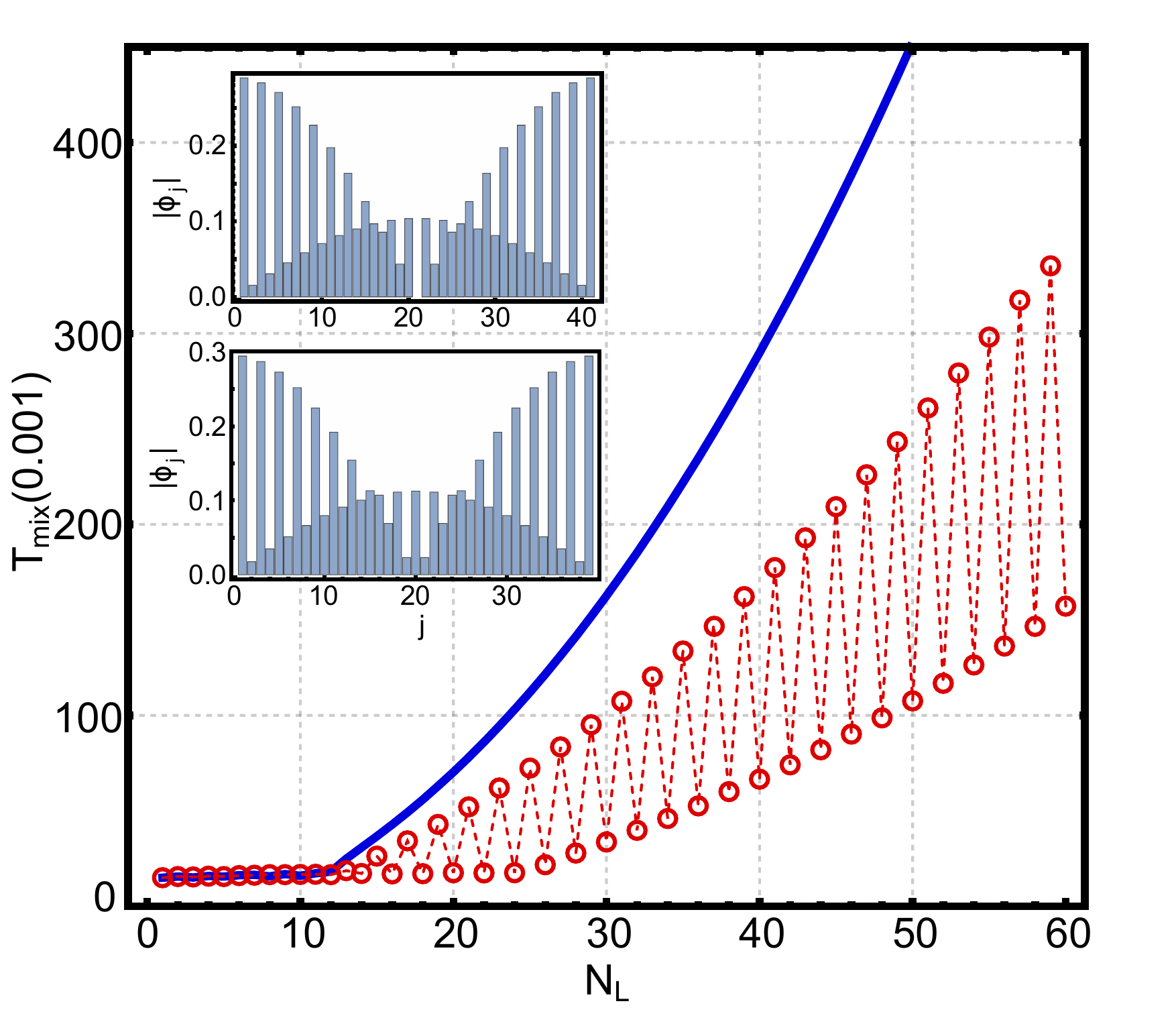} 
\end{center}
\caption{ Main panel shows mixing time $T_{mix}(0.001)$ in the units of $\Gamma^{-1}$ for different positions of the initially excited node in dependence on $N_L$. Solid line corresponds to the initially excited first node, circle marks show mixing time for the initially excited middle node (i.e., $N_L$-th node for the odd $N_L$ and $N_L+1$-th node for the even $N_L$ with total of $2N_L+1$ nodes). The lower inset shows absolute values of the eigenvector elements for the eigenvector corresponding to the  eigenvalue with the second smallest imaginary part for an odd $N_L=19$, the upper inset shows the same but for an even $N_L=20$. For all the panels, $v/\Gamma=3$.}
\label{fig4}
\end{figure}

The spectral features mentioned above directly influence the mixing behavior. Eq.(\ref{eig3}) shows that similarly to the DBS case, the mixing time at $v/\Gamma$ larger than the position of all the exceptional points depends mainly on the precision parameter $\epsilon$ (and, of course, the value of $\Gamma$). This can be seen with the estimation of $T_{mix}(\epsilon)$ shown in the main panel of Fig.\ref{fig2}(c), where the dependence of $T_{mix}(\epsilon)$ on the ratio $v/\Gamma$ is shown for different asymmetry angles $\phi$. The solid, dash-dotted, and dashed curves correspond to $\phi=0.25\pi, 0.23\pi, 0.1\pi$ with a single leftmost node initially excited, $\alpha_1(0)\neq 0$. The most immediate observation is that for the same single-node initial excitation, the symmetric structure mixes slower at $v/\Gamma$ larger than the exceptional point corresponding to the largest value of this ratio for $\phi=\pi/4$ (we call it "LREP"). Indeed, it can be seen from Eq.(\ref{eig3}) that before the LREP
\[\mathrm{min}\left\{{\rm Im}(\lambda^{\pm}_k(\phi\neq \pi/4))\right\}>
\mathrm{min}\left\{{\rm Im}(\lambda^{\pm}_k(\pi/4))\right\}\] 
for $\phi\in[0,\pi/2]$. 
As shown in Fig.\ref{fig2}(c), for strongly asymmetric case mixing might be orders of magnitude faster. However, in that case mixing looses its practical meaning. As it was pointed above, the mixed state is strongly localized.

Another curious albeit intuitive feature of asymmetric structures is the rather strong quantitative difference in $T_{mix}$ dependence on $v/\Gamma$ for different initial excitations. In Fig.\ref{fig2}(c), the dashed and dotted curves differ only by the initial excitation; $\alpha_1(0)\neq 0$ for the dashed curve and $\alpha_{N_L+1}(0)\neq 0$ for the dotted one with all the other nodes being initially of zero amplitudes. When the initial excitation is far from the region where the stationary distribution is localized, the mixing is slower.  

As we shall see below, the choice of initial excitation can strongly affect mixing even in the symmetric case.

\subsection{Mixing and chain length}
As we have seen in the previous Subsection, asymmetry does not introduce any qualitative changes in conventional mixing in RL lattices. So, for simplicity sake, in this Subsection we discuss conventional mixing for symmetric RL lattices in dependence on the number of nodes. 

Generally, it conforms to the patterns discussed in the Section IV. As it is obvious from the eigenvalues given by Eq.(\ref{eig3}), for any given $v/\Gamma$ the system will eventually cross the LREP as $N_L$ increases. Indeed, for the position of LREP from Eq. (\ref{eig3}), one has $v/\Gamma\approx N_L/\sqrt{2}\pi$ for $N_L\gg 1$.  Thus, with increasing $N_L$, the system eventually becomes susceptible to slowing of mixing (with $T_{mix}\propto N_L^2$, as for the usual Markovian chain mixing \cite{aldous}). 

The examples of such behaviour are shown in Fig.\ref{fig3} for only the edge node being initially excited. Here the dependence of $T_{mix}(0.001)$ on $N_L$ is shown for different ratios $v/\Gamma$. The solid, dash-dotted, and dashed lines correspond to $v/\Gamma=3,5,7$, respectively. It is seen that after crossing LREP the mixing time grows like $N_L^2$. The inset shows that before crossing the LREP, $T_{mix}$ grows much slower. Moreover, as discussed in Section IV, before the LREP the mixing time is rather close for different $v/\Gamma$.

Thus, we have the recipe for optimizing mixing time for the RL lattice of any given length: one needs just to increase the interaction strength between the nodes so that it crosses the LREP.

\begin{figure}[htb]
\begin{center}
\includegraphics[width=0.6\linewidth]{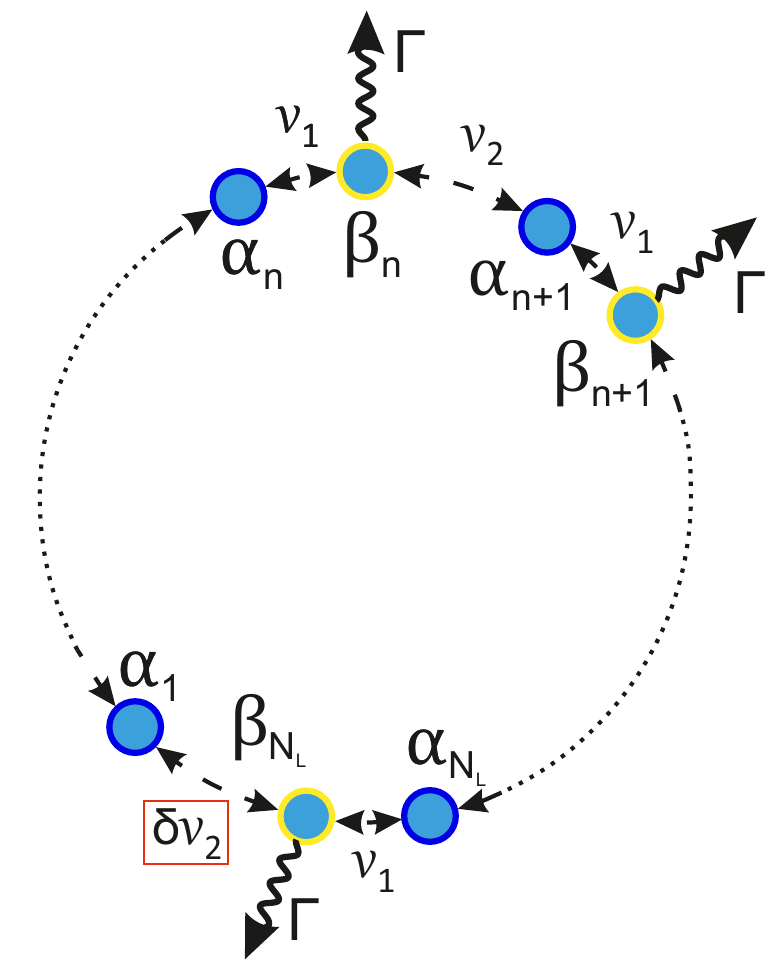} 
\end{center}
\caption{The scheme of the ring RL lattice. There are $N_L$ loss-free nodes $\alpha_k$ and $N_L$ dissipative nodes $\beta_k$ with loss rate $\Gamma$. The coupling constant of the nodes $\alpha_k$ and $\beta_k$ is $v_1$; the coupling constant of the nodes $\beta_k$ and $\alpha_{k+1}$ is $v_2$. The last dissipative node $\beta_{N_L}$ is coupled to the first node $\alpha_1$ with the interaction constant $\delta v_2$; $\delta$ is a balancing parameter.}
\label{fig5}
\end{figure}

\begin{figure}[htb]
\begin{center}
\includegraphics[width=0.9\linewidth]{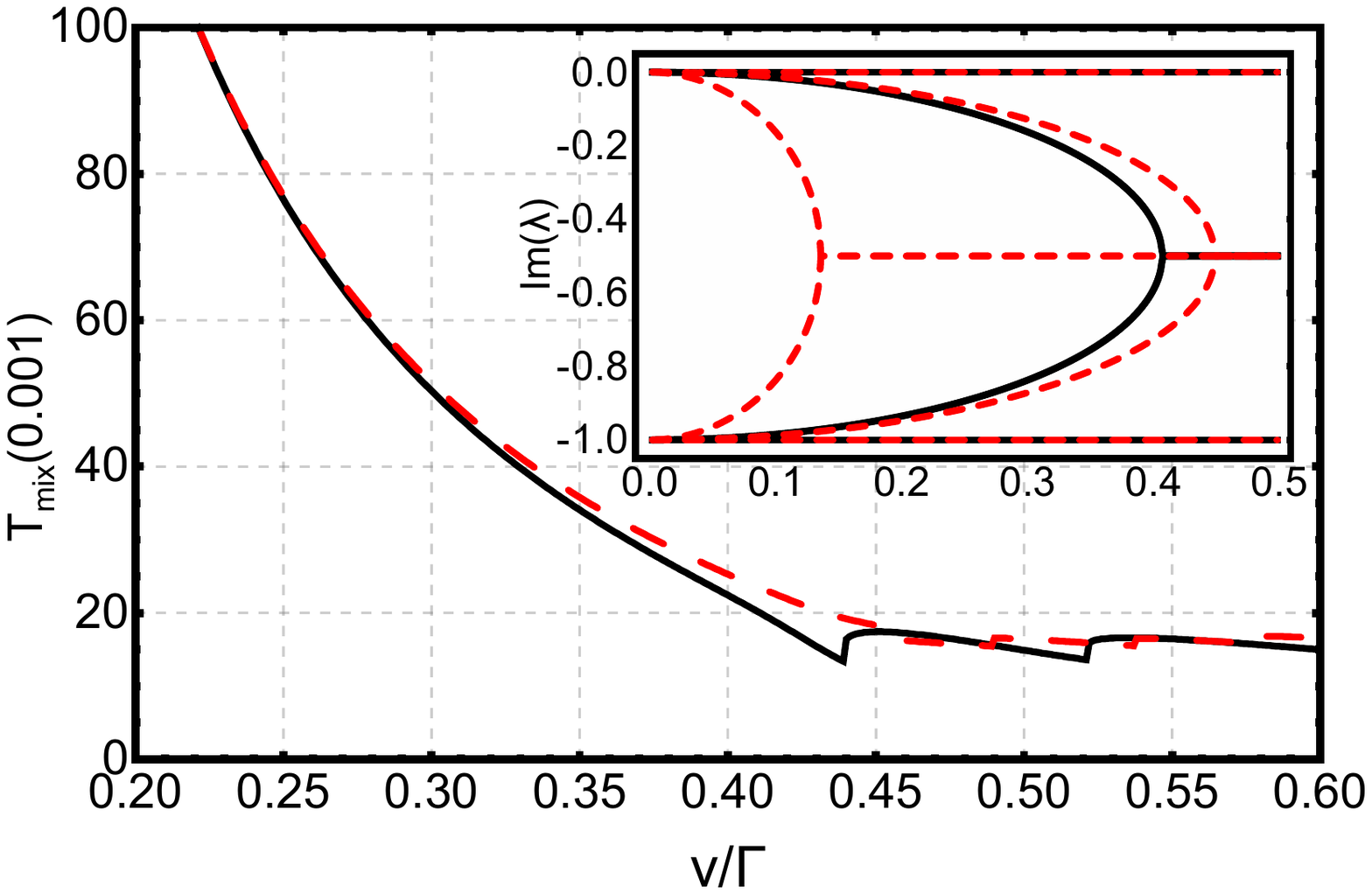} 
\end{center}
\caption{Mixing time as a function of $v/\Gamma$ for the ring RL lattice with $N_L=3$ in the units of $\Gamma^{-1}$ for the initially excited node $\alpha_1$ as given by Eq.(\ref{mix}) with $\epsilon=0.001$. Inset shows the imaginary parts of the eigenvalues of the Hamiltonian (\ref{ham4}) with the balancing parameter (\ref{cond1}) in dependence on the ratio $v/\Gamma$. For all the figure, the solid line corresponds to the symmetric structure, the dashed line corresponds to $\phi=0.15\pi$.}
\label{fig6}
\end{figure}

\subsection{Mixing and initial states}

The recipe of increasing the mixing rate described in the previous Subsection is not always available. One cannot increase the interaction strength indefinitely. For example, in the system based on single-mode waveguides, increasing interaction strength means decreasing the distance between the waveguides. This eventually breaks the weak-coupling and single-mode approximations \cite{sue77}. 

However, it appears that one can avoid crossing the LREP and reach the fast mixing regime by judicious choice of the initial state. Even for just a single initially excited node, its choice might mean a lot. Fig.\ref{fig4} shows that exciting initially a node in the middle of the lattice instead of the edge one, one can make mixing much faster. Moreover, the region of log-like fast mixing is much wider for even $N_L$ than for odd $N_L$ (the ratio is nearly two for the example shown in Fig.\ref{fig4}). The reason can be related to the structure of the eigenvectors. For even $N_L$, the eigenvector corresponding to the second smallest eigenvalue has zero element corresponding to the central node, $k=N_L+1$. The initial excitation of this node effectively shifts the LREP to the second largest exceptional point in Fig.\ref{fig2}. Thus, having just one initially excited node, one can drastically decrease the mixing time by making the RL lattice longer. 

Of course, it is possible to exploit the features of eigenstates in a similar manner making the initial excitation non-local for effectively shifting the LREP. Indeed, adjusting the amplitudes of initial excitation of $M+1$ nodes, outside of exceptional points one can always make the initial state orthogonal to any of $M$ eigenvectors extending the region of log-like mixing time dependence with the system's size. 

\begin{figure}[htb]
\begin{center}
\includegraphics[width=0.9\linewidth]{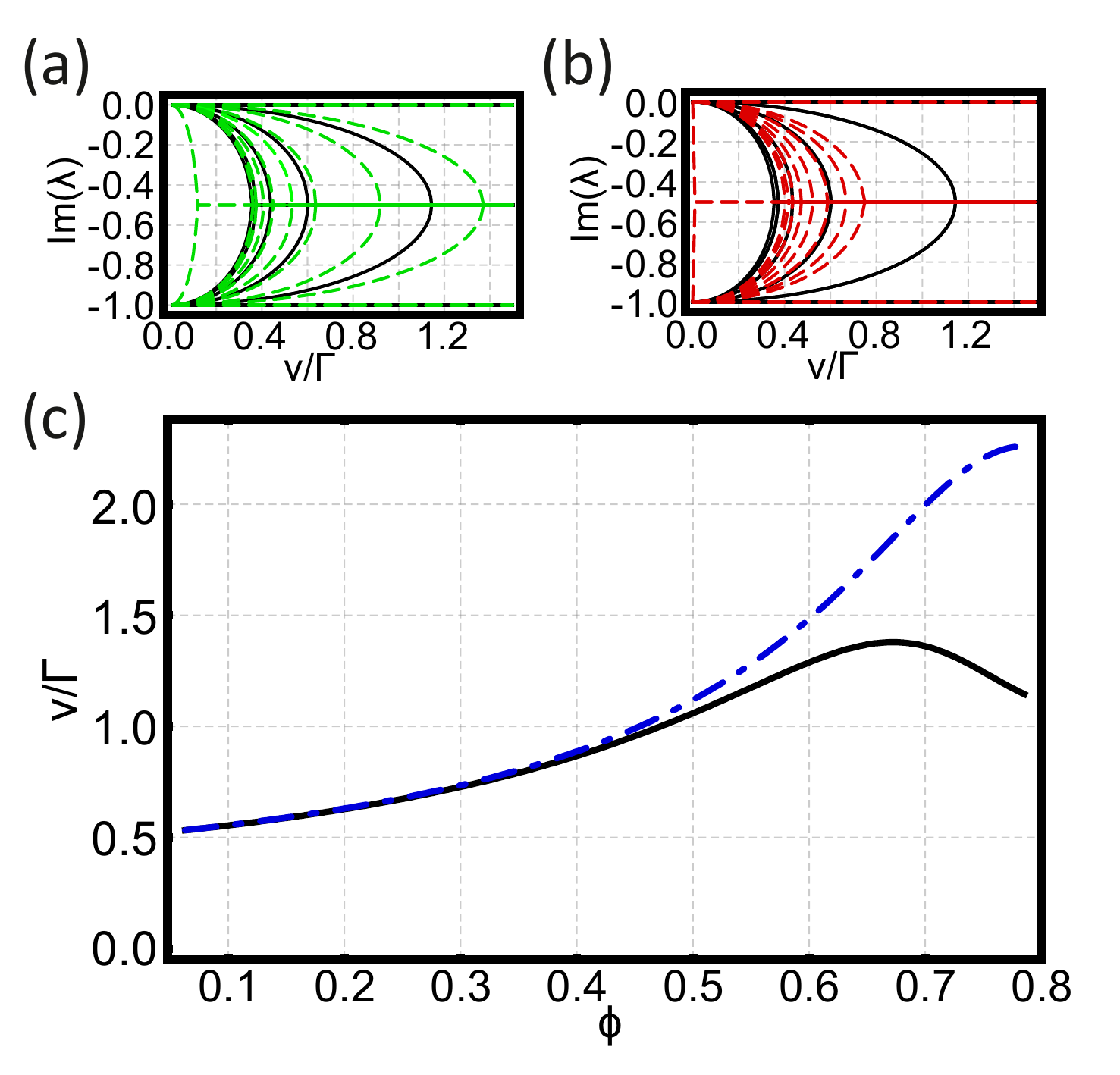} 
\end{center}
\caption{ (a,b) Imaginary parts of the eigenvalues of the Hamiltonian (\ref{ham4}) with the balancing parameter (\ref{cond1})  in dependence on the ratio $v/\Gamma$ for the RL ring with $N_L=10$. On both these panels the solid lines show $\mathrm{Im}(\lambda_k)$ for the symmetric case versus the ratio $v/\Gamma$. The dashed line in the panel (a) shows $\mathrm{Im}(\lambda_k)$ for the slightly asymmetric case, $\phi=0.22\pi$. The dashed line in the panel (b) shows $\mathrm{Im}(\lambda_k)$ for the strongly asymmetric case, $\phi=0.1\pi$. The panel (c) shows the position of the LREP  in dependence on the asymmetry angle $\phi$ for the linear structure (dash-dotted line) and the ring structure (solid line). For the linear structure $N_L=9$,  for the ring structure $N_L=10$.}
\label{fig7}
\end{figure}

\section{Ring RL lattice}

It is easy to surmise that optimization of the mixing time can be achieved also by changing geometry of the lattice. Here we demonstrate that a simple modification of RL lattice, namely, closing it into a ring structure, can bring considerable advantages in mixing. Such a ring structure can be readily realized, for example, by laser waveguide writing in a bulk dielectric \cite{Mukherjee2017DissipativelyCW}. A scheme of this modification is shown in Fig.\ref{fig5}. It coincides with the scheme shown in Fig.\ref{fig2}(a) with just one difference: there is an additional lossy node (say, $\beta_{N_L}$) coupling the nodes $\alpha_1$ and $\alpha_{N_L}$. Also, the coupling of this node with the node $\alpha_1$ is weighted by a parameter $\delta$. So, the Hamiltonian for the system shown in Fig.\ref{fig5} is given by 
\begin{eqnarray}
\label{ham4}
\nonumber
H_{2n,2n}=-i\Gamma, \quad H_{2n-1,2n}=H_{2n,2n-1}=v_1, \\
H_{2n,2n+1}=H_{2n+1,2n}=v_2, \quad n=1,2\ldots N_L-1, \\
\nonumber
H_{2N_L-1,2N_L}=H_{2N_L,2N_L-1}=v_1, \\
\nonumber
H_{2N_L,2N_L}=-i\Gamma, \quad H_{1,2N_L}=H_{2N_L,1}=\delta v_2.
\end{eqnarray}
The meaning of the balance parameter $\delta$ can be easily inferred from the Hamiltonian (\ref{ham4}). To have a non-vacuum stationary state and a possibility of conventional mixing, one needs to satisfy
\begin{equation}
v_1^{N_L}+(-1)^{N_L+1}\delta v_2^{N_L}=0.
\label{cond1}
\end{equation}
In the case of a symmetric structure, it leads to $\delta=1$ for even $N_L$ and $\delta=-1$ for odd $N_L$. Further, we consider only the conventional mixing with RL rings balanced as given by Eq.(\ref{cond1}).  
\begin{figure}[htb]
\begin{center}
\includegraphics[width=0.9\linewidth]{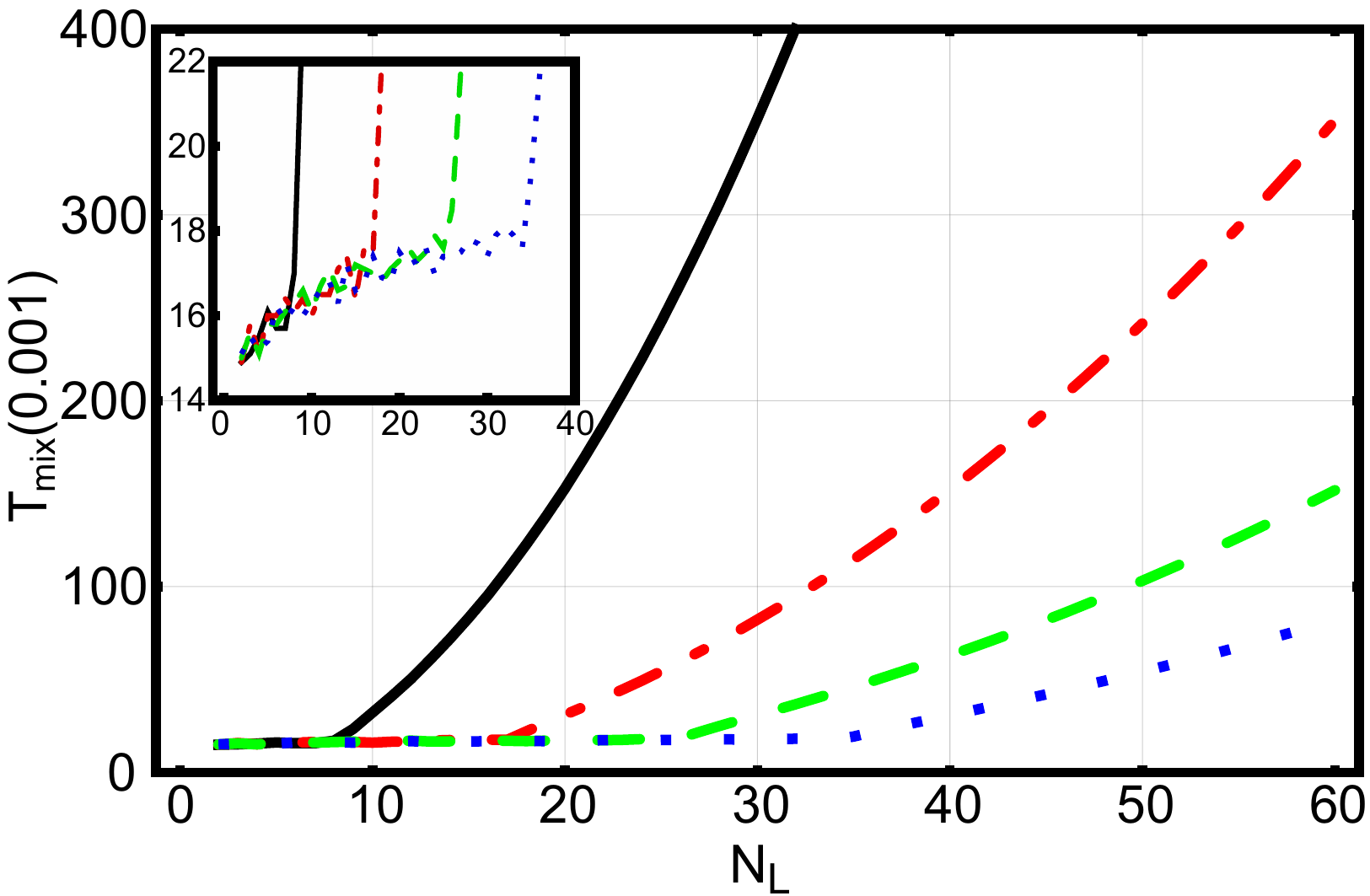} 
\end{center}
\caption{Dependence of the mixing time $T_{mix}$ on the number of lossy nodes $N_L$ for the ring RL lattice. {In the main panel, the solid and dash-dotted lines correspond to $v/\Gamma=1,2$ to the symmetric structure.} The dash-dotted line corresponds to $v/\Gamma=2$, $\phi=0.22\pi$; the dotted line corresponds to $v/\Gamma=2$, $\phi=0.2\pi$. The inset shows $T_{mix}$ for smaller vertical scale; here solid, dash-dotted, dashed and dotted lines correspond to $v/\Gamma=1,2,3,4$. For all the curves, $\epsilon=0.001$. Initially only the first lossless node is excited.}
\label{fig8}
\end{figure}

\subsection{The simplest ring structure}

To clarify how the mixing on a RL ring occurs, let us consider the simplest ring structure with non-trivial non-vacuum stationary state. It is the ring with three non-lossy nodes, $N_L=3$. As shown in the inset of Fig.\ref{fig6}, for the symmetric structure, the imaginary parts of the eigenvalues of the Hamiltonian (\ref{ham4}) behave quite similarly to the ones for the DBS considered in Section III. There is also just a single exceptional point. As the main panel of Fig.\ref{fig6} shows, the mixing time quickly grows for values of $v/\Gamma$ below this point. However, there are two significant differences. Firstly, there is an eigenvalue independent on $v$ and equal to $\Gamma$. 

Secondly and rather curiously, here we have convalescence of two diabolical points (more exactly, diabolical branches \cite{doi:10.1098/rspa.1984.0022}) into the exceptional points.  The upper and lower solid curves in the inset in Fig.\ref{fig6} show the degenerate eigenvalues corresponding to pairs of different eigenvectors.   These features are easily seen from the Hamiltonian (\ref{ham4}). Indeed, one can get from Eq.(\ref{ham4}) the following equations for the compound amplitudes of dissipative nodes  $B_1=\beta_1-\beta_3$, $B_2=\beta_1+2\beta_2+\beta_3$, $B_3=\beta_2-\beta_1-\beta_3$: 
\begin{eqnarray}
\label{b2eqs}
\nonumber
\frac{d^2}{dt^2}B_{1,2}+\Gamma\frac{d}{dt}B_{1,2}+\frac{3}{2}v^2B_{1,2}=0, \\
\frac{d^2}{dt^2}B_3+\Gamma\frac{d}{dt}B_{3}=0.
\end{eqnarray}
Curiously, as Eqs.(\ref{b2eqs}) show, the exceptional  point retains its diabolic character: it is actually a pair of exceptional points with different eigenvectors. 
 
Notice that the superposition $B_3$ corresponds to the unconventionally mixed state. Also notice that the symmetric ring structure gives the exceptional point at the interaction constant $\sqrt{3/2}$ times smaller that in the case of the DBS (\ref{eig1}). The ring structure provides faster mixing than the linear DBS for the same $v$ values.

However, symmetry breaking in the case of the ring RL lattice leads to the consequences opposite to those of the DBS. Asymmetric coupling leads to lifting the degeneracy (see the dashed lines in the inset in Fig.\ref{fig6} corresponding to $\phi=0.15\pi$), so that the LREP shifts to the larger ratios $v/\Gamma$. As expected, this results in slower mixing as seen in the main panel of Fig.\ref{fig6} (see the dashed line corresponding to $\phi=0.15\pi$). Another difference from the DBS is the independence of $T_{mix}$ of the initially excited lossless node position in the asymmetric case. 

\begin{figure}[htb]
\begin{center}
\includegraphics[width=0.9\linewidth]{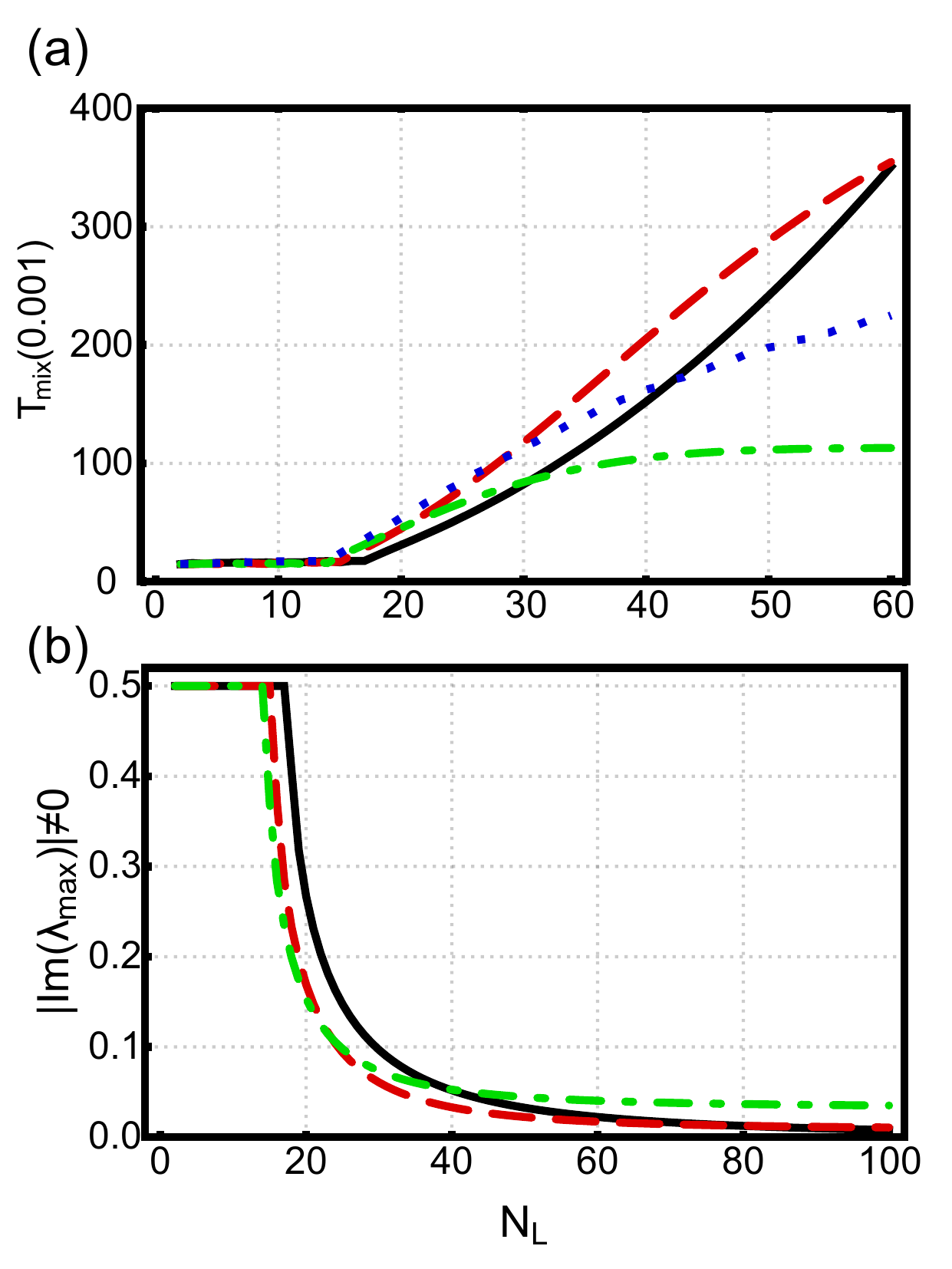} 
\end{center}
\caption{(a) Dependence of the mixing time $T_{mix}$ on the number of lossy nodes $N_L$ for the asymmetric ring RL lattice in comparison with the symmetric RL structure; for all the curves $\epsilon=0.001$, $v/\Gamma=2$. Solid, dashed and dash-dotted curves correspond to $\phi=0.25\pi,0.24\pi,0.23\pi$ and initially excited ${N_L}$-th lossless node; dotted curve corresponds to $\phi=0.23\pi$ and excitation of the lossless node opposite to the ${N_L}$-th lossless node. (b) Dependence of the second smallest $|Im(\lambda)|$ of the number of nodes $N_L$ for $v/\Gamma=2$. Solid, dashed and dash-dotted curves correspond to $\phi=0.25\pi,0.24\pi,0.23\pi$.}
\label{fig9}
\end{figure}

\subsection{Mixing and the ring perimeter}

Now let us discuss how the mixing time depends on the ring perimeter which is twice the number of dissipative or lossless nodes. Just like in the previous subsection, we are discussing here only the balanced rings allowing for conventional mixing. First of all, one needs to notice that the balanced ring RL lattice shows dissipation rates spectra quite similar to those for the linear RL lattice (see Fig.\ref{fig7}(a,b)). One also has a set of exceptional points. With increasing of the ratio $v/\Gamma$ the system eventually crosses the LREP. However, there are also some rather remarkable differences apart from the already noticed presence of the unconventionally mixed state corresponding to $\mathrm{Im}(\lambda)=-\Gamma$. For the symmetric case and for odd $N_L$, all the exceptional points are twice degenerate. For even $N_L$, the exceptional point corresponding to the lowest ratio is not degenerate (for that reason one sees just five exceptional points for solid lines in Fig.\ref{fig7}(a,b)). Making the lattice asymmetric (i.e., for $\phi\neq\pi/4$) removes degeneracy. As one can see in Fig.\ref{fig7}(c) (dash-dotted line), for the linear RL lattice, increase in asymmetry resulted in monotonous shift of the LREP to the region of lower $v/\Gamma$. This is not the case for the ring structure. If the mixing angle decreases from the symmetric value $\pi/4$, the LREP shifts to the larger values of $v/\Gamma$ (see the dashed line in Fig.\ref{fig7}(a)) corresponding to $\phi=0.22\pi$). But further decrease of $\phi$ leads to the reverse motion of the LREP which eventually shifts to lower $v/\Gamma$ than for the symmetric RL ring (see the dashed line in Fig.\ref{fig7}(b) corresponding to $\phi=0.1\pi$ and the solid line in Fig.\ref{fig7}(c) showing how the LREP moves with the asymmetry angle $\phi$ for the ring structure).  

Another important difference from the linear RL lattice is that for the asymmetric ring structure, the exceptional point shifts to the lower ratios $v/\Gamma$ with decreasing $\phi$ with no limit similar to those described by Eq.(\ref{eig3}). So, asymmetry can break the "dissipative coupling" approximation which is always valid for the linear RL structure for sufficiently low ratios $v/\Gamma$ regardless of the asymmetry.   

The specific features of the spectra for the ring structures significantly affect the mixing. As one can see in Fig.\ref{fig8}, for the symmetric structure the scaling is just like as for the linear RL lattice. There is the log-like scaling for the structure below the LREP and $N_L^2$ scaling above it. Similar to the smaller ring structure discussed in the previous subsection, the $N_L^2$ scaling starts at considerably larger values of $N_L$ for the ring structure than for the linear structure with the same $v/\Gamma$ (for example, one needs nearly twice as large $N_L$ for $v/\Gamma=3$, as follows from the comparison between Figs.\ref{fig3} and \ref{fig8}). So, the symmetric balanced ring structure is considerably better for a mixing device than a linear structure.

Much larger difference from the linear case appears for the balanced asymmetric structure. One can see from Eq.(\ref{cond1}) that the balancing parameter $\delta$ for an asymmetric structure grows with the number of nodes. This induces quite a remarkable spectral feature: stabilization of the second smallest dissipation rate with the increasing number of nodes, $N_L$. An example of such a spectral curiosity is shown in Fig.\ref{fig9}(b) for $v/\Gamma=2$ and asymmetry angles $\phi=0.24\pi$ (dashed curve) and $\phi=0.23\pi$ (dash-dotted curve). For comparison, the solid curve shows how the the second smallest dissipation rate tends to zero with increasing $N_L$ for the symmetric structure. Naturally, stabilization  of $Im(\lambda)$ induces returning to log-like scaling of the mixing time for asymmetric rings. Examples of such fast mixing restoration are shown in Fig.\ref{fig9}(a) for $v/\Gamma=2$ and asymmetry angles $\phi=0.24\pi$ (dashed curve) and $\phi=0.23\pi$ (dash-dotted and dotted curves). The dotted curve demonstrates qualitative dependence of the mixing time on the choice of the initial state; the initially excited lossless node is maximally distant from the $N_L$-th lossless node. 

Finally, it is worth reminding that asymmetry makes the mixed distribution localized. In particular, for the balanced ring structure shown in Fig.\ref{fig5}, the mixed state localizes near $\alpha_{N_L}$ like $[\cot\phi]^m$, $m$ being a number of lossless nodes counter-clockwise from the $N_L$-th lossless node. So, even despite restoration of the fast scaling,  asymmetric ring is hardly feasible as a practical mixing device.

\section{Conclusions}

In this work, we have demonstrated how one can optimize mixing in linear and circular Rudner-Levitov lattices \cite{PhysRevLett.102.065703}, which  are the Su–Schrieffer–Heeger lattices with the second sub-lattice being lossy \cite{PhysRevLett.42.1698}). The RL lattices have a number of important applications in photonics and quantum optics, so that the mixing optimization in such structures is quite important for the practical photonic devices realizing such structures. We have obtained a number of important results. First of all, we have demonstrated that for the finite RL lattice one can always reach a region of fast, log-like dependence of the mixing time of the number of the lattice nodes just by engineering the coupling between nodes and loss rates of the lossy nodes. The key to such a design is the position of the exceptional points of the lattice. For any fixed ratio of the interaction constant and the loss rate, as the number of nodes is increased, the system eventually crosses the last exceptional point and the mixing time acquires square dependence on the number of nodes. However, one can extend the region of log-like dependence by judicious choice of the initial state keeping only a few nodes initially excited. We have also shown that the ring RL lattice provides the wider region of log-like dependence than the linear RL lattice for the same ratios of interaction constant and loss rate and the same initial states. Also, we have shown that asymmetry of the balanced ring structure can restore log-like scaling for an arbitrary number of nodes. 

\section*{Acknowledgments}
I. P. and D. M acknowledge financial support from the The Belarusian Republican Foundation for Fundamental Research, grant F22B-008.

\end{document}